\documentclass{appolb}
\usepackage{graphicx}

\begin{document}
\title{The ReBB model at 8 TeV:\\
Odderon exchange is not a probability, but a certainty%
\thanks{Presented at ``Diffraction and Low-$x$ 2022'', Corigliano Calabro, Italy, Sept. 
2022.}%
}
\author{István Szanyi$^{1,~2,~3,~\dag}$, Tamás Csörgő$^{2,~3,~\ddag}$
\address{ $^1$E\"otv\"os University, H - 1117 Budapest, P\'azm\'any P. s. 1/A, Hungary;\\
$^2$Wigner FK, H-1525 Budapest 114, POB 49, Hungary;\\
$^3$MATE Institute of Technology,  K\'aroly R\'obert Campus, H-3200 Gy\"ongy\"os, M\'atrai \'ut 36, Hungary;\\
$^\dag$iszanyi@cern.ch \\
$^\ddag$tcsorgo@cern.ch}
}
\maketitle
\begin{abstract}
The Real Extended Bialas-Bzdak (ReBB) model study is extended to the 8 TeV $pp$ TOTEM elastic differential cross section data. The analysis shows that the ReBB model describes the $pp$ and $p\bar{p}$ differential cross section data in the limited $0.37 \leq -t \leq 1.2$ GeV$\null^2$ and $1.96 \leq \sqrt{s} \leq 8$ TeV kinematic region, in a statistically acceptable manner. In this kinematic region a greater than 30 $\sigma$ model-dependent Odderon signal is observed by comparing the $pp$ and ReBB extrapolated $p\bar{p}$ differential cross sections. Thus, in practical terms, within the framework of the ReBB model, Odderon exchange is not a probability, but a certainty.
\end{abstract}
  
\section{Introduction}

In a recent paper \cite{Csorgo:2020wmw}, published in July 2021, we showed that the Real Extended $p=(q,d)$ version of the Bialas-Bzdak (ReBB) model developed in Ref.~\cite{Nemes:2015iia} based on the original papers, Refs.~\cite{Bialas:2006kw,Bialas:2006qf}, and later improvements, Refs.~\cite{Nemes:2012cp,CsorgO:2013kua}, describes in a statistically acceptable manner the proton-proton ($pp$) and proton-antiproton ($p\bar{p}$) scattering data in the kinematic range of $0.546\leq\sqrt{s}\leq 7$ TeV and $0.37\leq -t\leq1.2$ GeV$^2$. With these results at hand, we reported an at least 7.08 $\sigma$,  discovery level Odderon effect\footnote{The ReBB model is based on R. J. Glauber's multiple diffraction theory, so it operates directly on the level of the elastic scattering amplitude of $pp$ and $p\bar p$ collisions. We obtain the C-even (Pomeron) and C-odd (Odderon) components of the elastic scattering amplitude  as the average and the difference of elastic proton-antiproton and proton-proton amplitudes.  For the details see  Appendix C of Ref.~\cite{Csorgo:2020wmw}.} by comparing the $pp$ and $p\bar{p}$ differential cross sections at the same energies utilizing model-dependent extrapolations of the differential cross-sections of elastic $pp$ scattering to $\sqrt{s}$ $=$ 1.96 TeV and elastic $p\bar p$ scattering up to the lowest measured energy at LHC, 2.76 TeV.
Extrapolating the $p\bar{p}$ scattering up to 7 TeV, the statistical significance of Odderon exchange increased to greater than 10 $\sigma$, however, in Ref.~\cite{Csorgo:2020wmw} this significance was not quantified more precisely, due to numerical limitations of CERN's Root, MS Excel, Wolfram Mathematica and similar data analysis software tools.

Based on our recently published paper~\cite{Szanyi:2022ezh}, we present here the results of the extension of the ReBB analysis to the new 8 TeV $pp$ differential cross section data of TOTEM, Ref.~\cite{TOTEM:2021imi}. 
In this Ref.~\cite{Szanyi:2022ezh} we also more precisely quantified the high significance of the Odderon observation by introducing an analytical approximation scheme (see the Appendix of Ref.~\cite{Szanyi:2022ezh}). 


\section{ReBB model and Odderon exchange at 8 TeV}\label{sec:rebb_8_tev}

\begin{figure}[!hbt]
 \centerline{
 \includegraphics[width=0.5\textwidth]{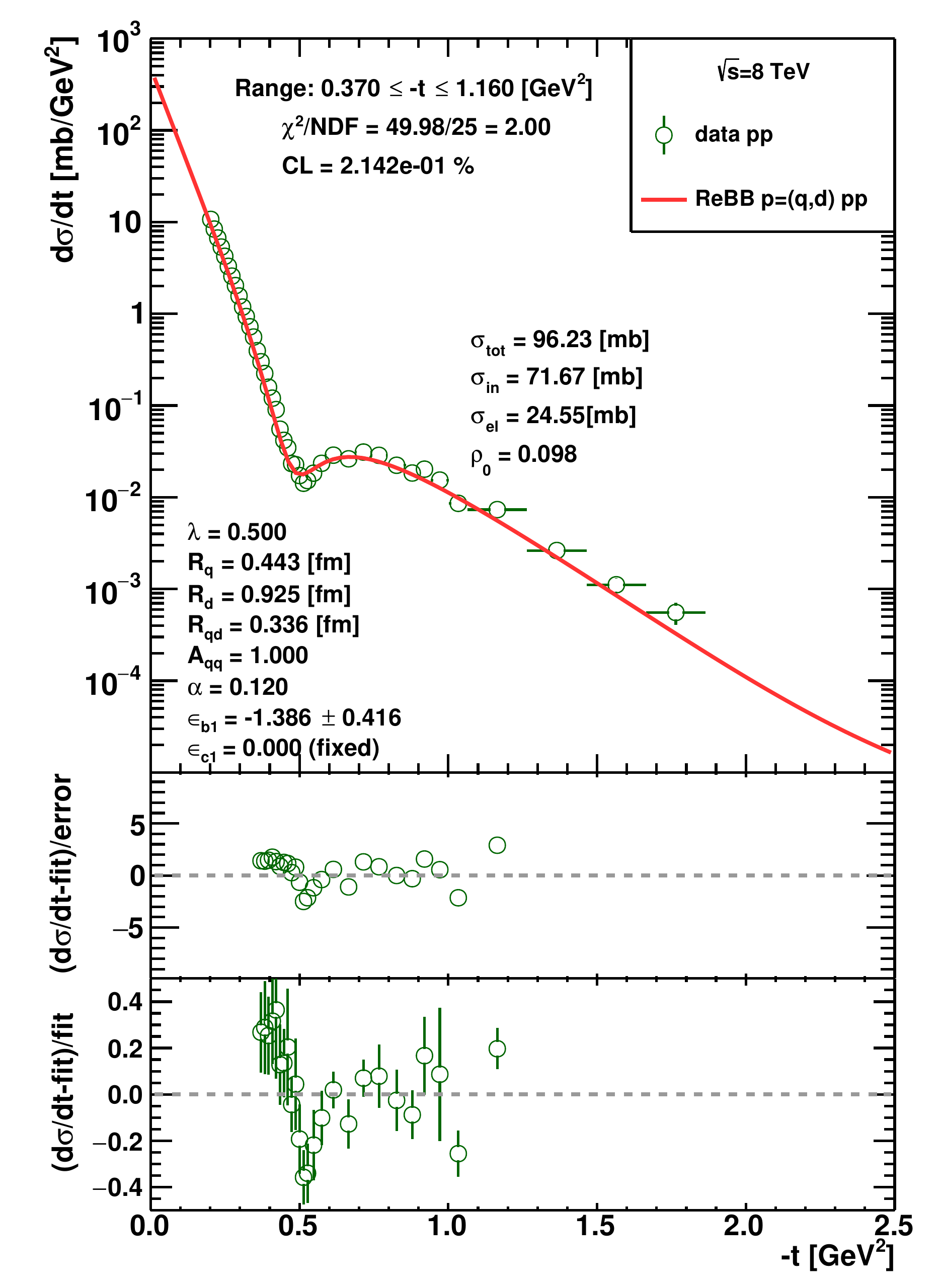}
 }
\caption{ 
Comparison of the $pp$ differential cross sections from Ref.~\cite{Szanyi:2022ezh}. The ReBB model calculations for $pp$ are based on Ref.~\cite{Csorgo:2020wmw} and  they agree, at 0.2 \% CL, with the recently published TOTEM $pp$ data at $\sqrt{s} = $ 8 TeV~\cite{TOTEM:2021imi}. 
}
\label{fig:H(x)_Odderon_1}
\end{figure}

As an extension to Ref.~\cite{Csorgo:2020wmw}, in Fig.~\ref{fig:H(x)_Odderon_1} we show the comparison of the $pp$ differential cross section calculated from the ReBB model --- using the energy calibration of the fit parameters done in Ref.~\cite{Csorgo:2020wmw} --- with the {\it final} 8 TeV $pp$ differential cross section data measured by TOTEM and published recently in Ref.~\cite{TOTEM:2021imi}. One can see that the energy-calibrated model, in its validity range, $0.37\leq -t \leq 1.2$ GeV$^2$, describes the data in a statistically acceptable manner, with a confidence level of 0.2 \%.

The ReBB model thus describes the data at 8 TeV in a limited kinematic region which is suitable to perform a search for Odderon exchange. As detailed and utilized recently in Refs.~\cite{Csorgo:2020wmw,Csorgo:2019ewn, TOTEM:2020zzr} a possible difference between $pp$ and $p\bar{p}$ measurable quantities at the TeV energy scale theoretically can be attributed only to the effect of a $t$-channel C-odd Odderon exchange. 

\begin{figure}[hbt!]
 \centerline{
 \includegraphics[width=0.55\textwidth]{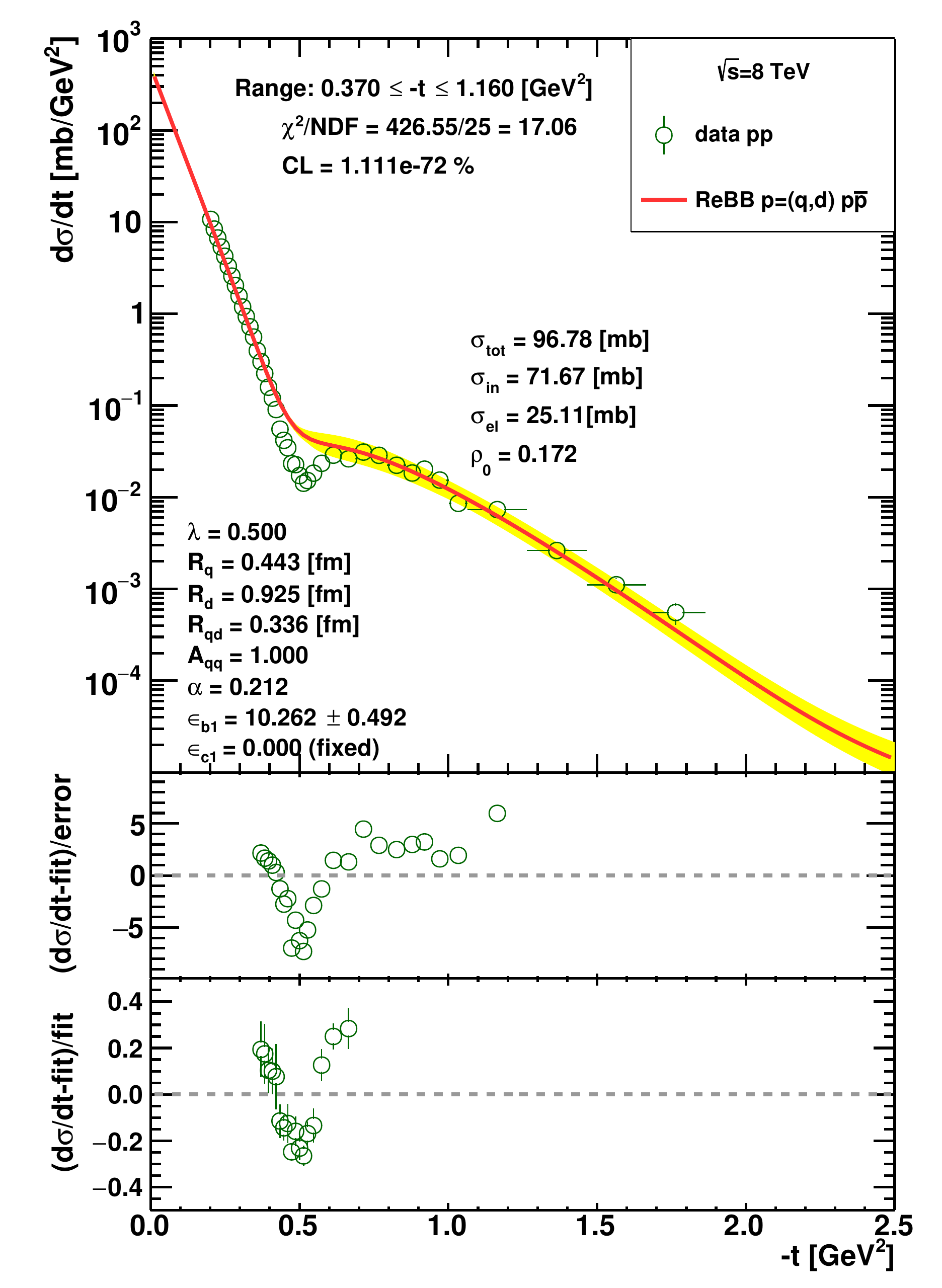}
 }
\caption{Comparison of the $p\bar{p}$ differential cross sections from Ref.~\cite{Szanyi:2022ezh}. The ReBB model calculations for $p\bar p$ are based on Ref.~\cite{Csorgo:2020wmw} and  they disagree, at $1.1\times 10^{-72}$ \% CL, with the recently published TOTEM $pp$ data at $\sqrt{s} = $ 8 TeV~\cite{TOTEM:2021imi}. }
\label{fig:H(x)_Odderon_3}
\end{figure}

The comparison of the $p\bar{p}$ differential cross section calculated from the ReBB model --- using the energy calibration of the fit parameters done in Ref.~\cite{Csorgo:2020wmw} --- with the 8 TeV $pp$ differential cross section data measured by TOTEM \cite{TOTEM:2021imi} is shown in Fig.~\ref{fig:H(x)_Odderon_3}, which indicates a difference between the $pp$ and $p\bar{p}$ differential cross sections with a probability of essentially 1, corresponding to a $CL=1-1.111\times10^{-74}$, \textit{i.e.}, an Odderon observation with a statistical significance $\geq$18.28 $\sigma$ (for the details of the significance calculation see the Appendix of Ref.~\cite{Szanyi:2022ezh}.).

The fits and the model-data comparisons are done by utilizing the $\chi^2$ definition developed by the PHENIX collaboration. This method is equivalent with the diagonalization of the covariance matrix if the experimental errors are separated into three different types: point-to-point fluctuating uncorrelated statistical and systematic errors (type A), point-to-point varying and 100\% correlated systematic errors (type B), and point-independent, overall correlated systematic uncertainties (type C). In our study, the available experimental errors of the analysed data can be and are categorized into these three types: horizontal and vertical $t$-dependent statistical errors (type A), horizontal and vertical $t$-dependent systematic errors (type B), and overall normalization uncertainties (type C).

The PHENIX method is validated by evaluating the $\chi^2$ from a full covariance matrix fit of the $\sqrt{s}$ = 13 TeV TOTEM differential cross-section data using the Lévy expansion method of Ref.~\cite{Csorgo:2018uyp}. The PHENIX method and the fit with the full covariance matrix result in the same minimum within one standard deviation of the fit parameters.
Thus the PHENIX method is a reasonable choice at energies, where the full covariance matrixes are not published. The exact form of the $\chi^2$ definition \footnote{The $\chi^2$ parameters $\epsilon_B$ and $\epsilon_C$ were considered as fit parameters in Ref.~\cite{Csorgo:2020wmw}, decreasing NDF. However $\epsilon_B$ and $\epsilon_C$, in fact, have a known central value (0) and a known standard deviation (1), hence they must be considered not only as fit parameters, but also new data points. Thus in the end they are not effecting the NDF. This was done in Ref.~\cite{Szanyi:2022ezh}, but this correction does not effect the conclusions drawn in Ref.~\cite{Csorgo:2020wmw}.} used in this analysis with correlation parameters, $\epsilon_B$ and $\epsilon_C$ resulting from such a classification of measurement errors can be found in Ref.~\cite{Csorgo:2020wmw}.

\begin{table*}[!hbt]
    \centering
    \begin{tabular}{cccccc}
        $\sqrt{s}$ (TeV) & $\chi^2$ & $NDF$ & CL & significance ($\sigma$)   \\ \hline 
        1.96 & 24.283 & 14 & 0.0423 & 2.0  \\ 
        2.76 & 100.347 & 22 & 5.6093 $\times10^{-12}$ & 6.8  \\ 
        7 & 2811.46 & 58 & $\textless$ 7.2853$\times 10^{-312} $ & $\textgreater$37.7  \\ 
        8 & 426.553 & 25 & 1.1111$\times 10^{-74}$ & $\geq$18.2   \\ 
    \end{tabular}
    \caption{Summary on Odderon signal observation significances in the ReBB model analysis from Ref.~\cite{Szanyi:2022ezh}. The significances higher than 8 $\sigma$ were calculated by utilizing an analytical approximation schema, detailed in the Appendix of the same paper~\cite{Szanyi:2022ezh}.}\label{tab:odsum}
\end{table*}

\section{ReBB model and Odderon at the TeV energy range}\label{sec:ReBBO}

Table~\ref{tab:odsum} summarises all the Odderon signal observation significances in our ReBB model analysis. The dataset at 7 TeV carries the largest, dominant Odderon signal, greater than 37.75 $\sigma$. The existence of a significant Odderon signal is confirmed with the new TOTEM data at 8 TeV, which provides an also clear-cut, greater than 18.28 $\sigma$ Odderon signal. The significance of the Odderon signal in the $\sqrt{s} = 2.76$ TeV TOTEM data is  $6.8$ $\sigma$. Within the framework of the ReBB model, no statistically significant Odderon signal is observed from the comparison of the $\sqrt{s} = 1.96$ TeV D0 data with ReBB model extrapolated elastic $pp$ differential cross-sections. 

Given that the datasets are independent measurements, we can evaluate their combined significances step by step, by adding  the individual $\chi^2$ and the individual $NDF$ values. Another option for combining the significances is Stouffer's method ($i.e.$ by summing the significances and dividing the sum by the square root of the number of summed significances) as used by TOTEM in Ref.~\cite{TOTEM:2020zzr}.  As it is detailed in Ref.~\cite{Szanyi:2022ezh} in Table 2, independently which method is used, the combination of the results at the two lowest energies, $i.e.$ 1.96 and 2.76 TeV, gives greater than 6 $\sigma$ significance for the Odderon exchange, while the combination of the results at $\sqrt{s} = $ 1.96, 2.76, 7 and 8 TeV gives a greater than 30 $\sigma$ significance.

\begin{figure}[hbt!]
	\centering
	\includegraphics[width=0.5\linewidth]{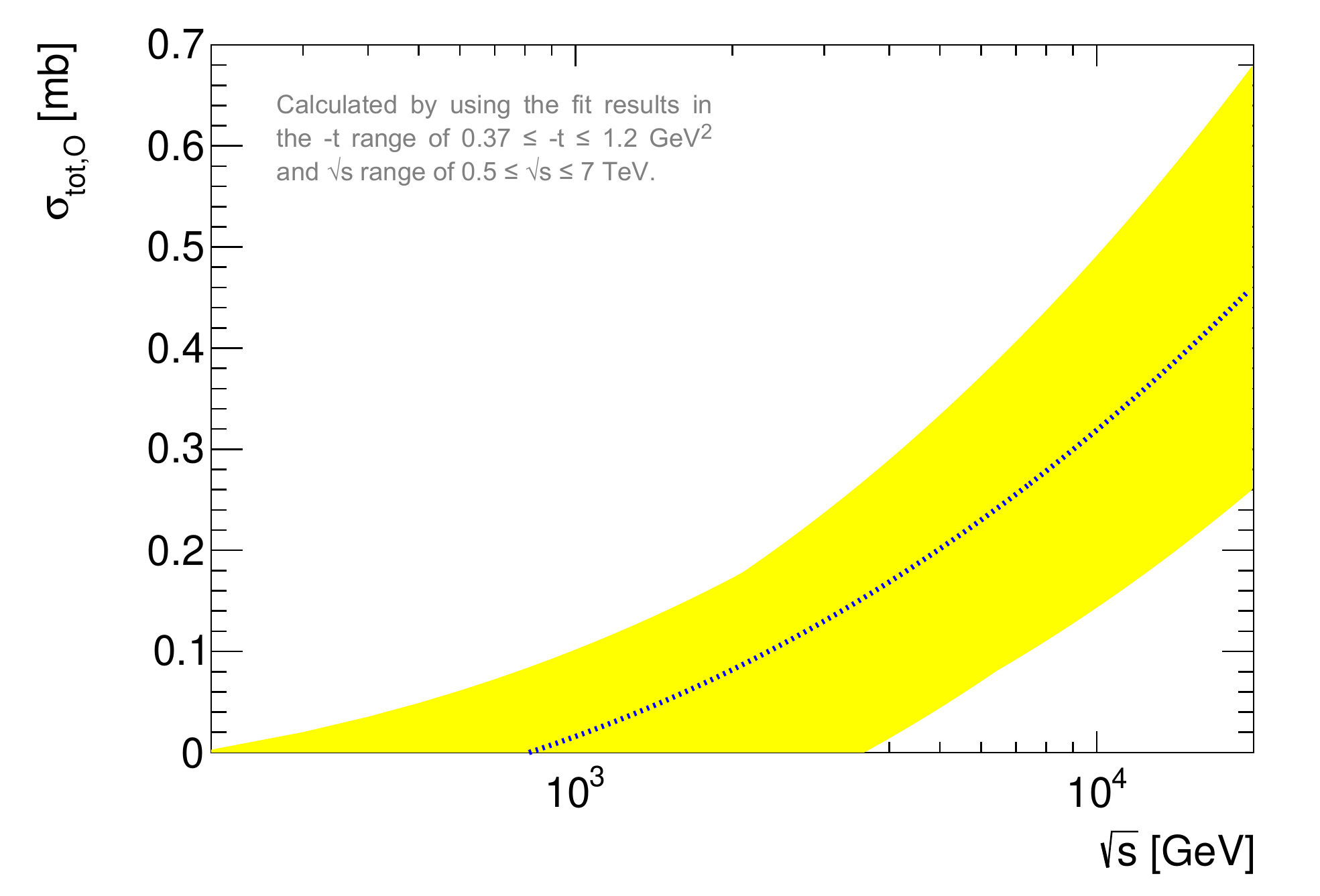}
	\caption{
        The Odderon total cross section, determined from the ReBB model in Ref.~\cite{Csorgo:2020wmw}, indicates a threshold effect for Odderon exchange. The Odderon contribution to the total cross-section starts to be statistically significant around 1 TeV.
	}
	\label{fig:sigmatot-Odderon}
\end{figure}

Fig.~\ref{fig:sigmatot-Odderon} shows the total cross-section (with systematic error band) obtained from the optical theorem using the ReBB model amplitude of Odderon exchange, as
evaluated from  the log-linear excitation functions of the model from Ref.~\cite{Csorgo:2020wmw}. The result indicates that the total cross-section of the Odderon exchange is sharply increasing in the few TeV energy range, but it is two orders of magnitude smaller than the contribution of the Pomeron exchange that is dominant at the same energy scale, as detailed in Ref.~\cite{Csorgo:2020wmw}.

\section{Summary}\label{sec:summ}

The Real Extended Bialas-Bzdak (ReBB) model describes all the available $pp$ and $p\bar{p}$ differential cross-section data in the kinematic range of $0.546\leq\sqrt{s}\leq 7$ TeV and $0.37\leq -t\leq1.2$ GeV$^2$ in a statistically acceptable  manner.
The statistical significance of  Odderon exchange is greater than 30 $\sigma$
when the results obtained from  $\sqrt{s} = $1.96, 2.76, 7, and 8 TeV are combined. Thus. within the framework of the ReBB model, Odderon exchange is not a probability, but a certainty at the TeV energy scale.

\section{Acknowledgments}\label{sec:summ}
We thank A. Papa and his team for their kind hospitality and for organizing an inspiring and useful meeting. Our research has been supported by the Hungarian NKFIH grant K133046 and  the ÚNKP-22-3 New National Excellence Program.
\bibliographystyle{unsrt}  
\bibliography{Odderon-Letter}

\end{document}